\documentclass[sigconf,nonacm]{acmart}

\AtBeginDocument{%
  }

\setcopyright{none}
\renewcommand\footnotetextcopyrightpermission[1]{}

\begin{document}

\title{The Unreasonable Effectiveness of Data for Recommender~Systems}

\author{Youssef Abdou}
\email{youssef.abdou@student.uni-siegen.de}
\affiliation{
  \institution{University of Siegen}
  \city{Siegen}
  \country{Germany}
}

\begin{abstract}
  In recommender systems, collecting, storing, and processing large-scale interaction data is increasingly costly in terms of time, energy, and computation, yet it remains unclear when additional data stops providing meaningful gains. This paper investigates how offline recommendation performance evolves as the size of the training dataset increases and whether a saturation point can be observed. We implemented a reproducible Python evaluation workflow with two established toolkits, \texttt{LensKit} and \texttt{RecBole}, included $11$ large public datasets with at least $7$ million interactions, and evaluated $10$ tool--algorithm combinations. Using absolute stratified user sampling, we trained models on nine sample sizes from $100{,}000$ to $100{,}000{,}000$ interactions and measured $NDCG@10$. Overall, raw $NDCG$ usually increased with sample size, with no observable saturation point. To make result groups comparable, we applied min--max normalization within each group, revealing a clear positive trend in which around $75\%$ of the points at the largest completed sample size also achieved the group's best observed performance. A late-stage slope analysis over the final $10$--$30\%$ of each group further supported this upward trend: the interquartile range remained entirely non-negative with a median near $1.0$. In summary, for traditional recommender systems on typical user--item interaction data, incorporating more training data remains primarily beneficial, while weaker scaling behavior is concentrated in atypical dataset cases and in the algorithmic outlier \texttt{RecBole BPR} under our setup.
\end{abstract}



\maketitle

\section{Introduction}
\subsection{Background}
Recommender systems rely on a limited number of parameters, with the dataset size being significant but practically controllable. This work investigates the relationship between dataset size and system performance to guide researchers and practitioners in making informed design choices aligned with their objectives.

\subsection{Research Problem}
Since large volumes of high-quality, usable data are difficult to gather, the value of relying on ever-larger datasets should be clearly justified. Researchers have highlighted challenges such as scalability, data sparsity, and privacy concerns across domains like e-commerce and media streaming \cite{raza2026}. At the same time, industrial recommender systems must balance potential performance gains against system cost and latency \cite{zou2025}.

Related evidence from practice suggests that non-algorithmic factors can dominate measured performance: changing the GUI has been shown to boost CTR by as much as $66\%$, potentially exceeding gains from algorithm tuning \cite{beel2021}. This influence is therefore a key confounder when conducting experiments on most standard-scale RecSys datasets. However, larger datasets can be more GUI-diverse and might therefore mitigate potential bias, since they are often accumulated over longer timelines during which the interface likely evolves.

Data pruning has on one hand been shown to result in up to a $3\%$ increase in performance when detrimental users are excluded \cite{meister2024}, while on the other hand a broader empirical study concluded that pruning should generally be avoided whenever possible \cite{beel2019}. Resolving this discrepancy in settings that incorporate larger data volumes remains an open question.

Model training, particularly on large datasets, is computationally expensive and resource-intensive. Our longest observed cases occurred among LensKit runs on \texttt{iPinYou} at the $1m$ scale, where every LensKit run required more than 20~hours of end-to-end runtime. We observed similarly long end-to-end runtimes across different libraries and datasets. Researchers describe deep learning model training as especially time-consuming on large datasets, which can take up to several months \cite{dong2020}.

Recent work by Vente et al. \cite{vente2024} emphasizes the considerable environmental cost of research on recommender systems, especially when using deep learning and large-scale datasets. Their analysis of full papers from the 2023 ACM Recommender Systems Conference shows that the average study consumed approximately 6,854 kWh of electricity, and that deep learning models required about eight times more energy than traditional machine learning methods without demonstrating clear performance benefits. The study further reports that the carbon footprint of recommender-system experiments has increased dramatically over the past decade, highlighting how larger datasets and more complex models intensify energy consumption and CO\textsubscript{2} emissions.

Consequently, the pursuit of larger datasets must be critically assessed; the environmental and computational costs require that the need for such data abundance be clearly justified.

\subsection{Research Question}
How does the performance of recommender systems evolve as the size of the training dataset increases? In particular, does performance improve steadily with more data or does it reach a point of diminishing returns? Is there an optimal dataset size at which a given recommender system likely achieves its best performance, beyond which additional training data yields no significant improvement? If such a saturation point exists, it needs to be identified, as recognizing it would greatly help reduce unnecessary computational costs and energy consumption.

\subsection{Research Hypothesis}
We hypothesize that, within recommender systems, predictive performance continues to improve as training data grows across the scales studied, rather than reaching a clear saturation point. This expectation is consistent with prior work showing that larger datasets often yield higher accuracy and more stable results \cite{ajiboye2015}, and with the broader view that scale can have an ``unreasonable'' positive effect \cite{halevy2009}. At the same time, earlier studies show that the apparent benefit of additional data can depend on the model and evaluation design \cite{catal2009,vabalas2019}. Determining whether saturation occurs is therefore crucial: evidence of saturation would shift attention toward understanding bottlenecks, whereas evidence against it would justify further investment in efficient large-scale data collection.

\section{Methodology}
\subsection{Overview}
We created a standard Python project that includes the libraries for two established tools, \texttt{LensKit} \cite{LKPY} and \texttt{RecBole} \cite{recbole}, as well as other necessary utility libraries. We gathered freely available datasets from a list \cite{recbole_datasets_page, recbole_datasets_repo} maintained by the \texttt{RecBole} team, prioritizing the larger options. An experimental run generally consists of multiple iterations as follows: first, load a prepared dataset; sample it to a specified number of rows using a chosen sampling strategy; select an algorithm from the list associated with the corresponding tool; provide the sampled dataset to the tool along with the selected algorithm; extract the resulting performance metric value; and finally, record it in the results file. The implementation and experiment scripts are publicly available \cite{repo}.

\subsection{Environment}
Python 3.11 and pinned library versions were required for compatibility across the workflow. These explicit versions are kept in the requirements file. The main utility libraries used were NumPy, pandas, and Matplotlib, while the selected recommender frameworks were \texttt{LensKit} 2025.2.0 and \texttt{RecBole} 1.2.1. The project is configured for SLURM-based \cite{slurm} execution and includes scripts for running jobs sequentially and in parallel, a Singularity \cite{singularity} container definition, and a Makefile \cite{make}. The OMNI HPC cluster of the University of Siegen was used for most computationally intensive jobs, where GPU-equipped nodes made the full experiment suite practical.

\subsection{Configurations}
Nearly all configurations are defined in the constants module. This includes settings shared between all \texttt{RecBole} algorithms, \texttt{RecBole} algorithm-specific settings, and \texttt{LensKit} settings. Hyperparameters and environment-related values are also handled in this module. In general, values were chosen to reduce runtime while remaining sufficient for proper training, since training computationally expensive algorithms on large datasets is not always practically feasible.

\subsection{Datasets}
Datasets from the list \cite{recbole_datasets_page, recbole_datasets_repo} previously mentioned in the overview subsection were selected with a clear preference for those with the largest interaction counts. Explicit interactions were generally preferred, since some algorithms do not work otherwise; therefore, most included datasets had a ratings column or an equivalent. All included datasets contained at least $7$ million interactions. Various forms of data cleaning were applied: the timestamp columns were uniformly removed and the uniqueness of the user ID–item ID pairs was enforced. Depending on the type of information, duplicates were either aggregated by summation or dropped so that only one record remained. In addition, dtypes were downcast to the lowest viable choice, and the datasets were stored as parquet \cite{parquet} files. Consequently, the loading time and the required storage capacity were considerably reduced.

\subsection{Sampling}
Differing sampling strategies were available as part of the experiment’s run mode. First, the size specification method (sizing, for simplicity) indicates whether the dataset is to be sampled to a fraction (percentage) or to an absolute number of rows. Second, the sample selection strategy can be one of four options: \texttt{Random Sampling}, \texttt{Stratified User Sampling}, \texttt{Stratified Item Sampling}, and \texttt{Stratified Hybrid Sampling}. \texttt{Random Sampling} is seeded and therefore deterministic, which suits the purpose of the experiment. All stratified variants aim to preserve representation ratios across groups, such that no group is over- or under-represented in the resulting sample. For example, in \texttt{Stratified User Sampling}, a user representing $10\%$ of the full dataset would also represent $10\%$ of the sample. The hybrid option combines the previous two approaches: stratified sampling by item, followed by stratified sampling by user. When targeting an absolute sample size with this strategy, fully preserving group ratios would typically yield a significantly larger sample. To address this, we slightly over-allocate group ratios and then trim the surplus via random sampling to obtain the exact desired size.

\subsection{Algorithms}
Five algorithms per tool were selected. The \texttt{LensKit} scoring models included were PopScorer, ItemKNNScorer, ImplicitMFScorer, BiasedMFScorer, and BiasedSVDScorer. For \texttt{RecBole}, the following recommendation models were selected: Pop, ItemKNN, BPR, NeuMF, and SimpleX.

\subsection{Experiment Variables}
Different values for experiment-related variables are directly configurable. The settings used to obtain our reported results were as follows: Top-10 recommendations for performance evaluation, absolute sizing, stratified user sampling, and nine sampling sizes ranging from $100$ thousand to $100$ million.

\section{Results}
\subsection{Terminology}
\paragraph{Instance.}
A result \emph{instance} is the output of a single evaluation run for a specific
$(\text{tool}, \text{algorithm}, \text{dataset}, \text{sample size})$ combination.
For example, \texttt{LensKit--Popularity--MovieLens-25M} $\mapsto \mathrm{NDCG} \approx 0.17$.
In Figures~\ref{fig:lines-common} and \ref{fig:lines-distinct}, each plotted point $(x,y)$ corresponds to one instance, where
$x$ is the sample size and $y$ is the $NDCG$ value.

\paragraph{Group.}
A result \emph{group} is the collection of instances that share the same
$(\text{tool}, \text{algorithm}, \text{dataset})$ combination, varying only in sample size.
Equivalently, a group is a set of points $\{(x_i, y_i)\}_{i=1}^n$ with fixed
tool/algorithm/dataset and different $x_i$.
In Figures~\ref{fig:lines-common} and \ref{fig:lines-distinct}, each colored line corresponds to one group.

\paragraph{Min--max normalization (within a group).}
For a given group $g$ with instances $\{(x_i, y_i)\}_{i=1}^n$, we normalize sample size and
$NDCG$ \emph{within that group} as
\[
x_i'=\frac{x_i-x_{\min}}{x_{\max}-x_{\min}},
\qquad
y_i'=\frac{y_i-y_{\min}}{y_{\max}-y_{\min}} .
\]
Here \(x_{\min}\) and \(x_{\max}\) are the minimum and maximum \(x\)-values within the group \(g\),
and \(y_{\min}\) and \(y_{\max}\) are the minimum and maximum \(y\)-values within the group \(g\).

\paragraph{Late-stage slope.}
For each min--max normalized group $g$, we extract a single \emph{late-stage slope} as follows.
Let $(x_2', y_2')$ be the point with maximal normalized sample size, i.e. $x_2' = 1$.
Choose $(x_1', y_1')$ as the point with the largest $x_1'$ such that
\[
0.1 \le x_2' - x_1' \le 0.3.
\]
The late-stage slope is then
\[
m_g = \frac{y_2' - y_1'}{x_2' - x_1'}.
\]

\begin{figure}[h]
  \centering
  \includegraphics[width=0.7\linewidth]{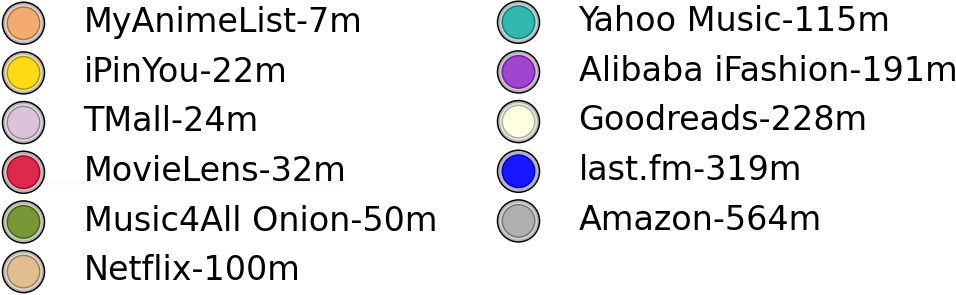}
  \caption{Legend}\label{fig:legend}
  \Description{Dataset coloring legend}
\end{figure}

\subsection{NDCG Trends}
\begin{figure}[h]
  \centering
  \includegraphics[width=\linewidth]{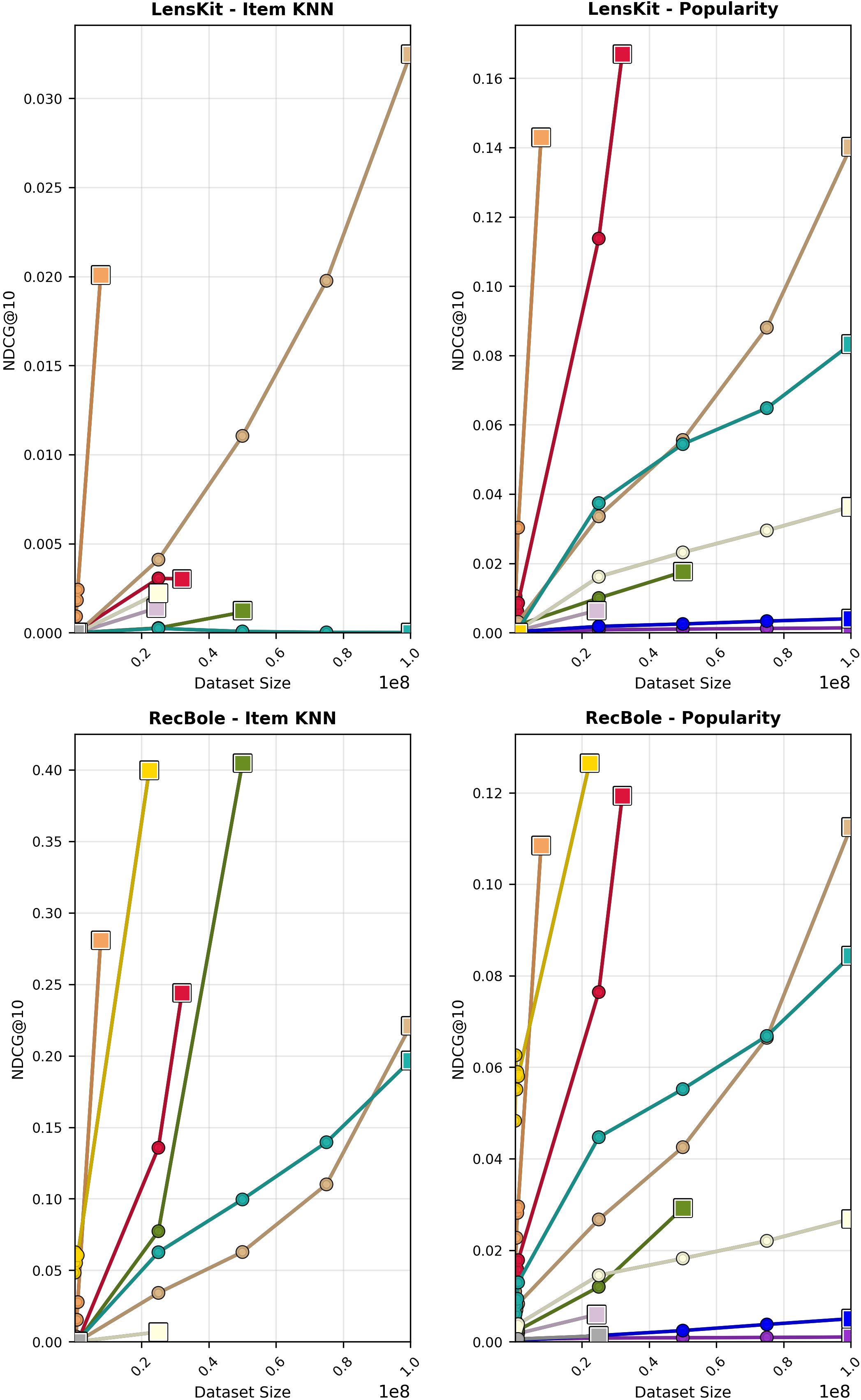}
  \caption{$NDCG@10$ vs. Sample Size}\label{fig:lines-common}
  \Description{Four line graphs the majority of their lines showing a clear positive trend}
\end{figure}
\begin{figure}[h]
  \centering
  \includegraphics[width=\linewidth]{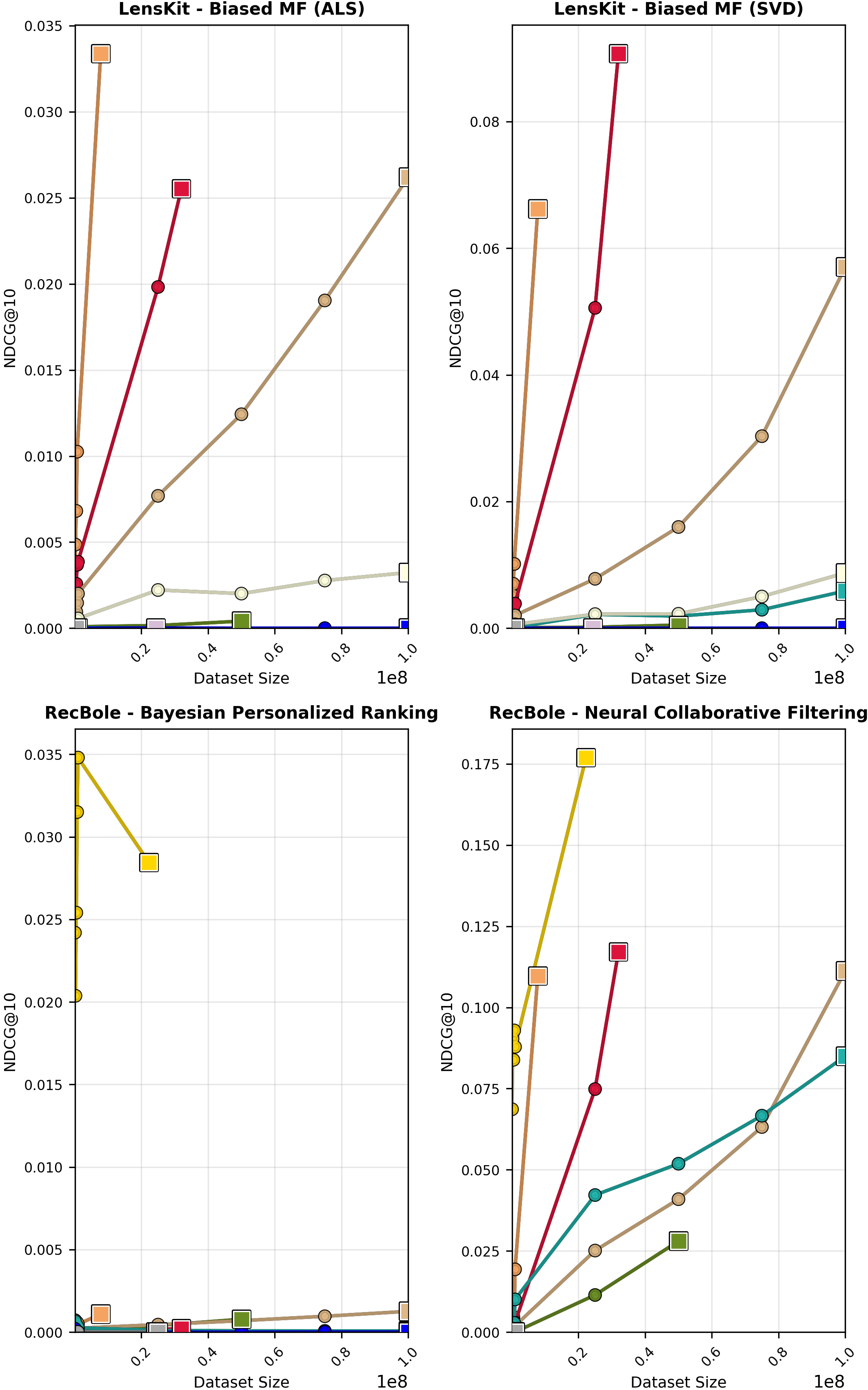}
  \caption{$NDCG@10$ vs. Sample Size cont.}\label{fig:lines-distinct}
  \Description{One out of the four graphs shows an outlying trend}
\end{figure}

\textbf{Performance increases steadily with sample size, with no visible saturation:} Figures \ref{fig:lines-common} and \ref{fig:lines-distinct} show the majority of the raw performance results. Each line plot shows $NDCG$ as a function of sampled input dataset size for the corresponding tool--algorithm pair. Within each plot, each colored line represents a result group. Each group contains instances at the sample sizes defined in the \emph{experiment variables}, up to either the full dataset size or the maximum configured sample size ($100m$), whichever is smaller. Some values could not be computed within a reasonable time and are therefore missing. In most plots, the overall trend increases with sample size, with no visible point at which $NDCG$ begins to diminish. Datasets that consistently show both a steep upward trend and a relatively high $NDCG$ include \texttt{MovieLens} and \texttt{Netflix}, reaching approximately $0.25$ at the full $32m$ and around $0.22$ at $100m$, respectively (both from \texttt{RecBole's Item KNN} results). In contrast, \texttt{Last.fm} shows no convincing performance improvement with increasing sample size in any group; its highest value is approximately $0.005$ when training on $100m$ rows using the \texttt{RecBole Popularity} model. \texttt{Alibaba-iFashion} shows even lower values, failing to exceed $0.002$ in all instances. Among the algorithms, \texttt{RecBole BPR} emerges as a clear outlier.

\subsection{Normalized Values Scatter}
\begin{figure}[h]
  \centering
  \includegraphics[width=\linewidth]{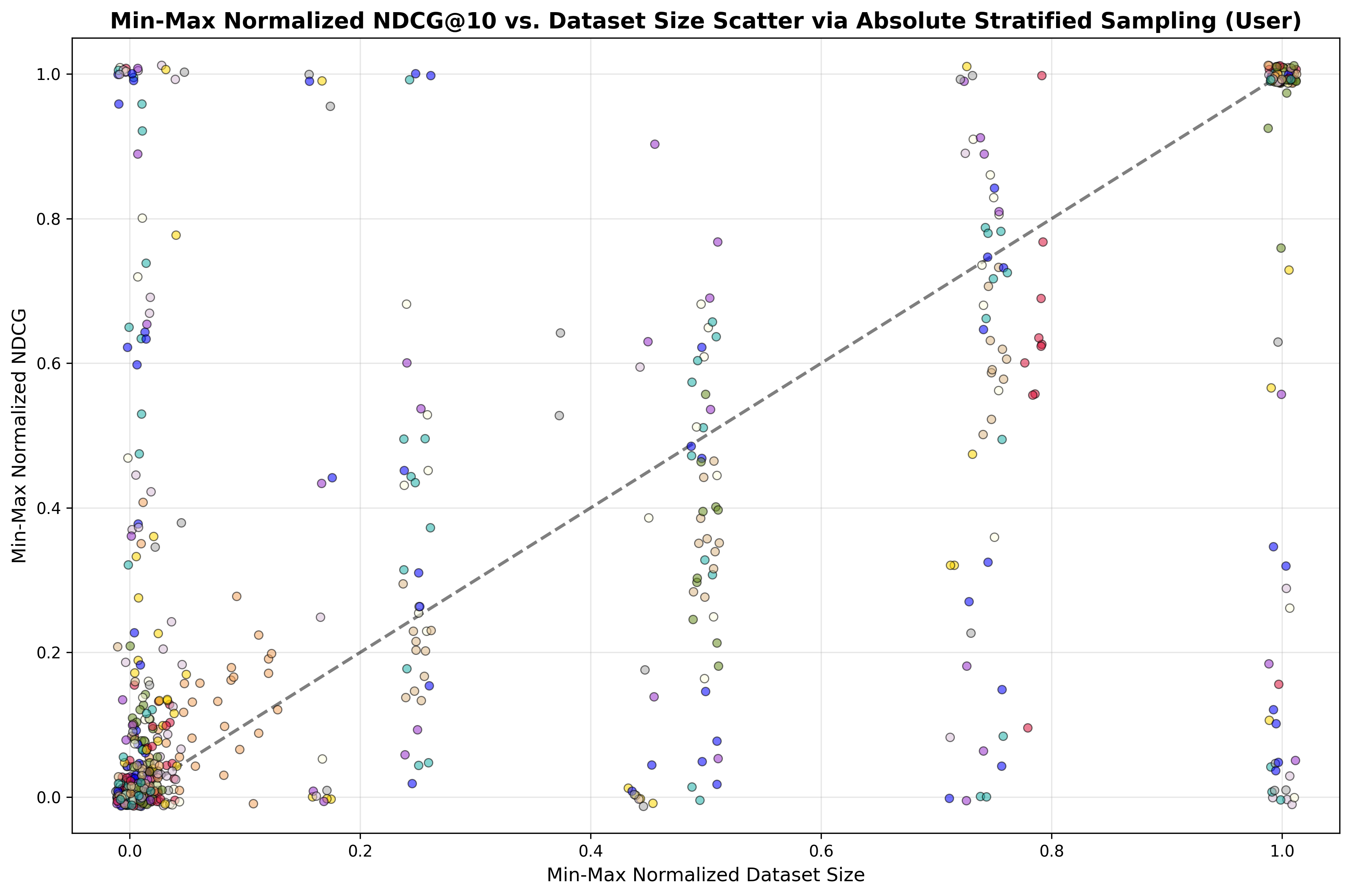}
  \caption{Normalized $NDCG@10$ vs. Sample Size Scatter}\label{fig:scatter}
  \Description{A Scatter plot more densely populated near the 0,0 point with a dashed diagonal reference line. The values slightly lag behind the reference but mimic the rising trend}
\end{figure}
\begin{figure}[h]
  \centering
  \includegraphics[width=\linewidth]{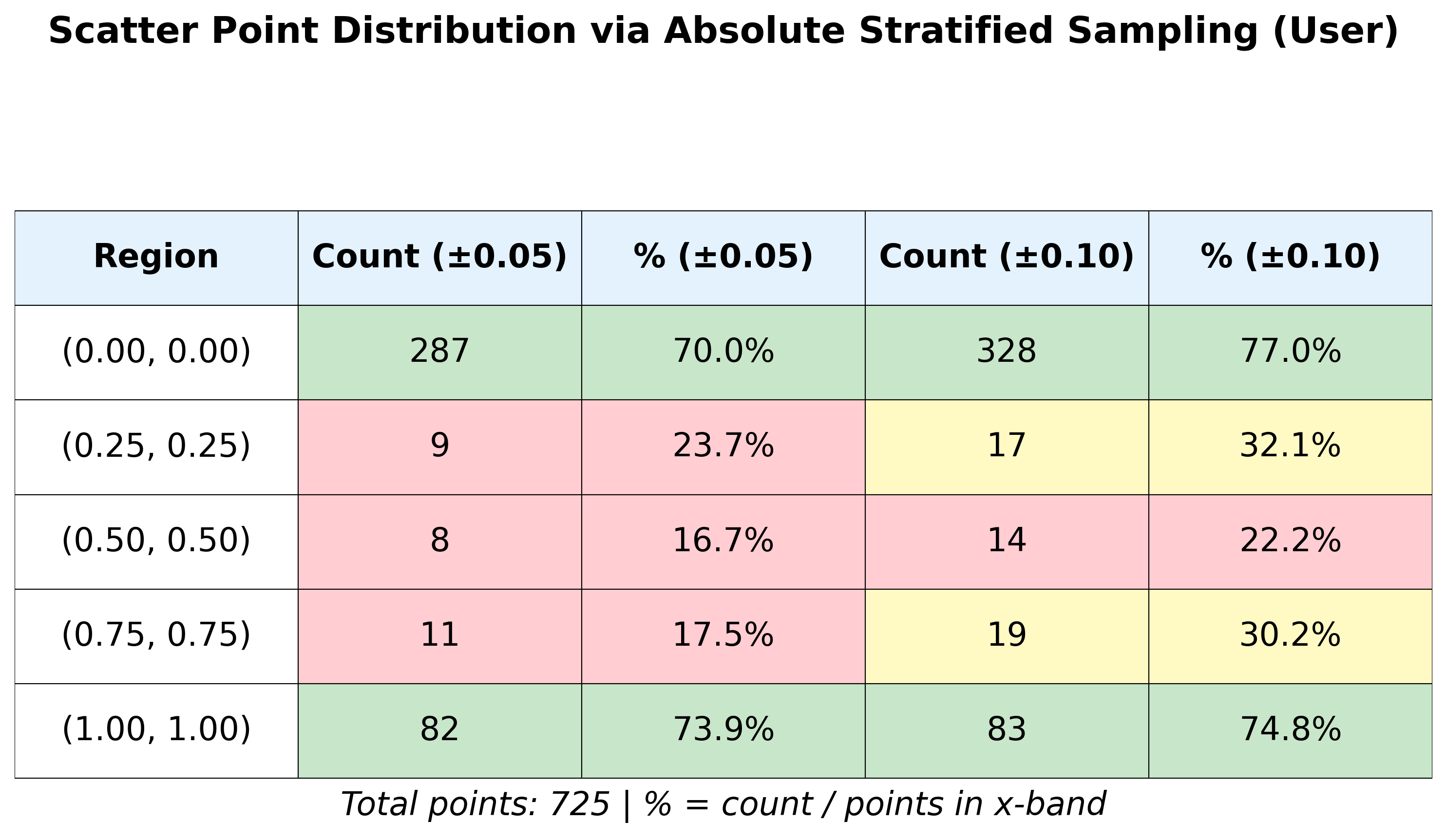}
  \caption{Figure \ref{fig:scatter} distributions metadata}\label{fig:scatter-metadata}
  \Description{A metadata table clarifying the data-point distributions}
\end{figure}

\textbf{Near-linear upward trend consistent across comparable normalized groups:} Figure~\ref{fig:scatter} reduces the gaps that arise when combining result groups with differing value ranges by applying per-group min--max normalization. Both axes are scaled to $[0.0, 1.0]$: on the $x$-axis, $1.0$ corresponds to the largest sample size within a group and $0.0$ to the smallest; the $y$-axis is scaled in the same way for $NDCG$. A point at $(0.0, 0.0)$ indicates that the lowest $NDCG$ in the group was obtained from the smallest sample size, whereas $(1.0, 1.0)$ indicates that the largest sample produced the highest $NDCG$. Colors distinguish datasets (not groups). For visual clarity, a uniform random jitter in $[-0.0125,0.0125]$ was applied, since overlap, especially at $(1.0, 1.0)$, was substantial; roughly $75\%$ of points with $x = 1.0$ also had $y = 1.0$, as per Figure~\ref{fig:scatter-metadata}. This suggests that, for most groups, peak performance occurs at the largest sample size. The region from $(0.0, 0.0)$ to $(0.2, 0.2)$ is densest because many of the sample sizes, defined in the \emph{experiment variables}, lie in the lower portion of each group's sample-size range (between $100k$ and $1m$), which therefore maps to small normalized $x$ values. The dashed diagonal is a reference line for a perfect positive linear relationship. Although clusters near both ends of the line are tightly concentrated, the sparser mid-range points align with the positive trend nonetheless. Outliers are visible across the plot, indicating that some result groups deviate from this positive pattern; however, this is consistent with the anomalous behaviors observed in the raw $NDCG$ trend plots shown in Figures~\ref{fig:lines-common} and \ref{fig:lines-distinct}.

\subsection{Late-Stage Slopes Distribution}
\begin{figure}[h]
  \centering
  \includegraphics[width=\linewidth]{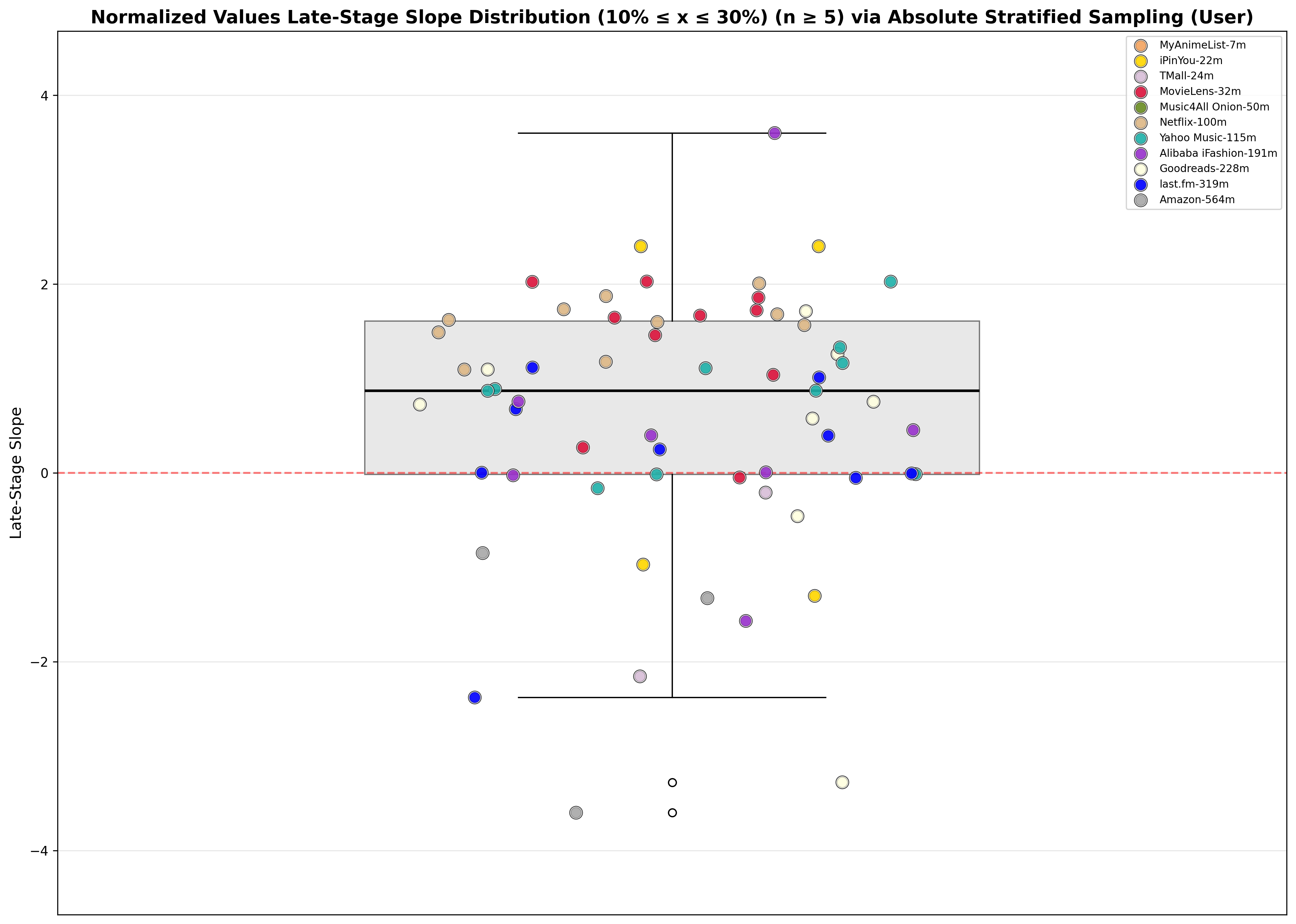}
  \caption{Normalized Values Late-Stage Slope Distribution}\label{fig:slopes}
  \Description{Vertical box-plot with a red horizontal reference line at value 0. the IQR span values 0 to 1.5}
\end{figure}
\textbf{Late-stage trends suggest increased performance if more data is introduced:} Figure~\ref{fig:slopes} shows a box plot that highlights the distribution of late-stage slope values for the normalized results shown in Figure~\ref{fig:scatter}, representing the "final rise" or the last observable trend. Each point is the slope over the final $x\%$ of the data, where approximately $10 \leq x \leq 30$. Result groups with fewer than five instances ($n < 5$) were excluded. The late-stage slope indicates whether a likely point of saturation has been reached, suggested by near-zero values that appear near the horizontal red reference line. Conversely, positive values above the reference suggest that adding more data will likely improve performance, pointing to a continued positive relationship that may reasonably be interpreted as truncated by data limitations. A small number of points (exactly $8$) lie on the reference line, with slopes approximately equal to $0.0$. The lower and upper whiskers are at around $-2.25$ and $3.5$, respectively, with two points falling below the lower whisker but none above the upper. The interquartile range spans around $0.0$ to approximately $1.5$, which is entirely non-negative. The reference line coincides with the lower quartile, and the median is nearly $1.0$.

\section{Conclusion}
This study supports the hypothesis that, for general recommender systems trained on standard numerical user--item feedback, $NDCG@10$ scales positively with additional interaction data rather than reaching a clear saturation point within the investigated range and evaluated offline setup. Across $11$ large public datasets, $10$ tool--algorithm combinations, and sample sizes ranging from $10^5$ to $10^8$ interactions, recommendation performance usually improves as more interaction data are provided. In many result groups, the largest completed sample yields the best observed $NDCG@10$, while late-stage trends remain predominantly positive. Weaker scaling behavior is concentrated in a limited set of atypical cases, most notably the datasets \texttt{Last.fm}, \texttt{Alibaba-iFashion}, and \texttt{Amazon}, and the algorithm \texttt{RecBole BPR}, where dataset-specific feedback characteristics or runtime-constrained configuration choices likely play a role. Taken together, these findings suggest that, for typical user--item interaction data and traditional recommender algorithms under the evaluated setup, additional training data often continues to deliver practically meaningful gains.

\bibliographystyle{ACM-Reference-Format}
\bibliography{references}

@article{ajiboye2015,
author = {Abdulraheem, Ajiboye and Abdullah Arshah, Ruzaini and Qin, Hongwu},
year = {2015},
month = {02},
pages = {75-84},
title = {Evaluating the Effect of Dataset Size on Predictive Model Using Supervised Learning Technique},
volume = {1},
journal = {International Journal of Software Engineering \& Computer Sciences (IJSECS)},
doi = {10.15282/ijsecs.1.2015.6.0006}
}

@article{halevy2009,
  title={The unreasonable effectiveness of data},
  author={Halevy, Alon and Norvig, Peter and Pereira, Fernando},
  journal={IEEE intelligent systems},
  volume={24},
  number={2},
  pages={8--12},
  year={2009},
  publisher={IEEE}
}

@article{catal2009,
  title={Investigating the effect of dataset size, metrics sets, and feature selection techniques on software fault prediction problem},
  author={Catal, Cagatay and Diri, Banu},
  journal={Information Sciences},
  volume={179},
  number={8},
  pages={1040--1058},
  year={2009},
  publisher={Elsevier}
}

@article{raza2026,
title = {A comprehensive review of recommender systems: Transitioning from theory to practice},
journal = {Computer Science Review},
volume = {59},
pages = {100849},
year = {2026},
issn = {1574-0137},
doi = {https://doi.org/10.1016/j.cosrev.2025.100849},
url = {https://www.sciencedirect.com/science/article/pii/S157401372500125X},
author = {Shaina Raza and Mizanur Rahman and Safiullah Kamawal and Armin Toroghi and Ananya Raval and Farshad Navah and Amirmohammad Kazemeini},
}

@inproceedings{recbole,
  author    = {Wayne Xin Zhao and Shanlei Mu and Yupeng Hou and Zihan Lin and Yushuo Chen and Xingyu Pan and Kaiyuan Li and Yujie Lu and Hui Wang and Changxin Tian and Yingqian Min and Zhichao Feng and Xinyan Fan and Xu Chen and Pengfei Wang and Wendi Ji and Yaliang Li and Xiaoling Wang and Ji{-}Rong Wen},
  title     = {RecBole: Towards a Unified, Comprehensive and Efficient Framework for Recommendation Algorithms},
  booktitle = {{CIKM}},
  pages     = {4653--4664},
  publisher = {{ACM}},
  year      = {2021}
}

@article{vabalas2019,
  title={Machine learning algorithm validation with a limited sample size},
  author={Vabalas, Andrius and Gowen, Emma and Poliakoff, Ellen and Casson, Alexander J},
  journal={PloS one},
  volume={14},
  number={11},
  pages={e0224365},
  year={2019},
  publisher={Public Library of Science San Francisco, CA USA}
}

@inproceedings{LKPY,
    title={LensKit for Python: Next-Generation Software for Recommender Systems Experiments},
    booktitle={Proceedings of the 29th ACM International Conference on Information and Knowledge Management},
    DOI={10.1145/3340531.3412778},
    author={Ekstrand, Michael D.},
    year={2020},
    month={Oct},
    extra={arXiv:1809.03125}
}

@article{singularity,
    doi = {10.1371/journal.pone.0177459},
    author = {Kurtzer, Gregory M. AND Sochat, Vanessa AND Bauer, Michael W.},
    journal = {PLOS ONE},
    publisher = {Public Library of Science},
    title = {Singularity: Scientific containers for mobility of compute},
    year = {2017},
    month = {05},
    volume = {12},
    url = {https://doi.org/10.1371/journal.pone.0177459},
    pages = {1-20},
    number = {5},
}

@misc{zou2025,
      title={A Survey of Real-World Recommender Systems: Challenges, Constraints, and Industrial Perspectives}, 
      author={Kuan Zou and Aixin Sun},
      year={2025},
      eprint={2509.06002},
      archivePrefix={arXiv},
      primaryClass={cs.IR},
      url={https://arxiv.org/abs/2509.06002}, 
}

@inproceedings{vente2024,
    author = {Vente, Tobias and Wegmeth, Lukas and Said, Alan and Beel, Joeran},
    title = {From Clicks to Carbon: The Environmental Toll of Recommender Systems},
    year = {2024},
    isbn = {9798400705052},
    publisher = {Association for Computing Machinery},
    address = {New York, NY, USA},
    url = {https://doi.org/10.1145/3640457.3688074},
    doi = {10.1145/3640457.3688074},
    booktitle = {Proceedings of the 18th ACM Conference on Recommender Systems},
    pages = {580–590},
    numpages = {11},
    location = {Bari, Italy},
    series = {RecSys '24}
}

@ARTICLE{dong2020,
  author={Dong, Qifei and Luo, Gang},
  journal={IEEE Access}, 
  title={Progress Indication for Deep Learning Model Training: A Feasibility Demonstration}, 
  year={2020},
  volume={8},
  number={},
  pages={79811-79843},
  doi={10.1109/ACCESS.2020.2989684}}

@inproceedings{beel2021,
    author = {Beel, Joeran and Dixon, Haley},
    title = {The ‘Unreasonable’ Effectiveness of Graphical User Interfaces for Recommender Systems},
    year = {2021},
    booktitle={Adjunct Proceedings of the 29th ACM Conference on User Modeling, Adaptation and Personalization},
    isbn = {9781450383677},
    publisher = {Association for Computing Machinery},
    address = {New York, NY, USA},
    url = {https://doi.org/10.1145/3450614.3461682},
    doi = {10.1145/3450614.3461682},
    pages = {22–28},
    numpages = {7},
    location = {Utrecht, Netherlands},
    series = {UMAP '21}
}

@inproceedings{meister2024,
  title={Removing Bad Influence: Identifying and Pruning Detrimental Users in Collaborative Filtering Recommender Systems.},
  author={Meister, Philipp and Wegmeth, Lukas and Vente, Tobias and Beel, Joeran},
  booktitle={RobustRecSys@ RecSys},
  pages={8--11},
  year={2024}
}

@inproceedings{beel2019,
  title={Data pruning in recommender systems research: Best-practice or malpractice},
  author={Beel, Joeran and Brunel, Victor},
  booktitle={13th ACM Conference on Recommender Systems (RecSys)},
  volume={2431},
  pages={26--30},
  year={2019},
  organization={CEUR-WS}
}

@misc{recbole_datasets_page,
  author       = {RUCAIBox},
  title        = {Dataset List | RecBole},
  howpublished = {\url{https://recbole.io/dataset_list.html}},
  year         = {2024},
  note         = {Accessed: 2026-03-08}
}

@misc{recbole_datasets_repo,
  author       = {RUCAIBox},
  title        = {RecSysDatasets: Public Data Sources for Recommender Systems},
  howpublished = {\url{https://github.com/RUCAIBox/RecSysDatasets}},
  year         = {2024},
  note         = {GitHub repository, accessed: 2026-03-08}
}

@misc{repo,
  author       = {Youssef Tarek Tewfik},
  title        = {unreasonable-effectiveness-recsys},
  howpublished = {\url{https://github.com/Youssef-Tarek-Tewfik/unreasonable-effectiveness-recsys}},
  year         = {2026},
  note         = {GitHub repository, accessed: 2026-03-19}
}

@inproceedings{slurm,
  author    = {Yoo, Andy B. and Jette, Morris A. and Grondona, Mark},
  title     = {SLURM: Simple Linux Utility for Resource Management},
  booktitle = {Job Scheduling Strategies for Parallel Processing},
  editor    = {Feitelson, Dror and Rudolph, Larry and Schwiegelshohn, Uwe},
  year      = {2003},
  publisher = {Springer Berlin Heidelberg},
  address   = {Berlin, Heidelberg},
  pages     = {44--60},
  isbn      = {978-3-540-39727-4}
}

@misc{parquet,
  title        = {Apache Parquet},
  author       = {{Apache Parquet contributors}},
  howpublished = {\url{https://parquet.apache.org/}},
  note         = {Accessed: 2026-03-17}
}

@article{make,
  author = {Feldman, Stuart I.},
  title = {Make — a program for maintaining computer programs},
  journal = {Software: Practice and Experience},
  volume = {9},
  number = {4},
  pages = {255-265},
  doi = {https://doi.org/10.1002/spe.4380090402},
  url = {https://onlinelibrary.wiley.com/doi/abs/10.1002/spe.4380090402},
  eprint = {https://onlinelibrary.wiley.com/doi/pdf/10.1002/spe.4380090402},
  year = {1979}
}

\end{document}